\documentclass[12pt]{article}

  \usepackage{graphicx}
  \usepackage{epsfig}
  \usepackage{amssymb}
  \usepackage{subfigure}

\evensidemargin - .5truecm
\begin{document}

\newcommand{\be}{\begin{equation}}
\newcommand{\ee}{\end{equation}}
\newcommand{\nn}{\nonumber}
\newcommand{\bea}{\begin{eqnarray}}
\newcommand{\eea}{\end{eqnarray}}
\newcommand{\bfig}{\begin{figure}}
\newcommand{\efig}{\end{figure}}
\newcommand{\bc}{\begin{center}}
\newcommand{\ec}{\end{center}}
\def\ad{\dot{\alpha}}
\def\ov{\overline}
\def\hlf{\frac{1}{2}}
\def\qrt{\frac{1}{4}}
\def\as{\alpha_s}
\def\at{\alpha_t}
\def\ab{\alpha_b}
\def\sq2{\sqrt{2}}
\newcommand{\smallz}{{\scriptscriptstyle Z}} %
\newcommand{\mz}{m_\smallz}
\newcommand{\smallw}{{\scriptscriptstyle W}}
\newcommand{\mw}{m_\smallw} 
\newcommand{\smallh}{{\scriptscriptstyle H}}
\newcommand{\mh}{m_\smallh}
\newcommand{\mt}{m_t}
\newcommand{\wh}{w_\smallh}
\def\th{t_\smallh}
\def\zh{z_\smallh}
\newcommand{\Mvariable}[1]{#1}
\newcommand{\Mfunction}[1]{#1}
\newcommand{\Muserfunction}[1]{#1}
%
%


\begin{titlepage}
\nopagebreak
{\flushright{
        \begin{minipage}{5cm}
         Rome1-1373/04 \\
         Freiburg-THEP 04/05 \\
         RM3-TH/04-6 \\
         IFUM-788/FT \\
        \end{minipage}        }

}
\renewcommand{\thefootnote}{\fnsymbol{footnote}}
\vskip 0.5cm
\begin{center}
\boldmath
{\Large\bf Two-loop light fermion contribution\\[3mm]
to Higgs production and decays}\unboldmath
\vskip 1.cm
{\large  U.~Aglietti 
\footnote{Email: Ugo.Aglietti@roma1.infn.it} ,}
\vskip .2cm
{\it Dipartimento di Fisica, Universit\`a di Roma ``La Sapienza'' and
INFN, Sezione di Roma, P.le Aldo Moro~2, I-00185 Rome, Italy} 
\vskip .2cm
{\large  R.~Bonciani 
\footnote{This work was supported  by the European Union under
contract HPRN-CT-2000-00149}
\footnote{Email: Roberto.Bonciani@physik.uni-freiburg.de} ,}
\vskip .2cm
{\it Fakult\"at f\"ur Mathematik und Physik, 
Albert-Ludwigs-Universit\"at
Freiburg, \\ D-79104 Freiburg, Germany} 
\vskip .2cm
{\large G.~Degrassi\footnote{Email: degrassi@fis.uniroma3.it}},
\vskip .2cm
{\it Dipartimento di Fisica, Universit\`a di Roma Tre and 
INFN, Sezione di Roma III, \\ Via della Vasca Navale~84, I-00146 Rome, Italy} 
\vskip .2cm
{\large A.~Vicini\footnote{Email: Alessandro.Vicini@mi.infn.it}}
\vskip .2cm
{\it Dipartimento di Fisica ``G. Occhialini'',
Universit\`a degli Studi di Milano and
INFN, Sezione di Milano,
Via Celoria 16, I--20133 Milano, Italy} 
\end{center}
\vskip 1.2cm

\begin{abstract}
We compute the electroweak corrections due to the light fermions 
to the production cross section $\sigma( g g \rightarrow H)$ and
to the partial decay widths $\Gamma \left( H \rightarrow  \gamma \,
\gamma  \right)$ and $\Gamma \left( H \rightarrow  g \, g \right)$. 
We present  analytic results for these corrections that are  expressed in 
terms of Generalized Harmonic Polylogarithms.
We  find that for the gluon fusion production cross section and for the 
decay width $\Gamma \left( H \rightarrow g g \right)$ the corrections 
are large in the Higgs mass region below 160 GeV where they reach up 
to 9\% of the lowest order term. For the decay width $\Gamma \left( H 
\rightarrow \gamma \gamma \right)$ the corrections for Higgs mass 
above 160 GeV can reach  $-$10\% of the lowest order term. 

\end{abstract}
\vfill
\end{titlepage}    
\setcounter{footnote}{0}
One of the great successes of the experimental program carried out at LEP
has been to put a firm lower bound on the Higgs mass, 
$\mh > 114$ GeV~\cite{unknown:2003ih}, and 
at the same time, together with the information coming from SLD, to
give a strong indirect evidence that the Higgs boson, the still missing 
particle of the Standard Model (SM), should be relatively light with a high
probability for its mass to be below 200 GeV. The search for the Higgs
boson is one of the main objective of the Tevatron and the future
Large Hadron Collider (LHC), that are supposed to span all the Higgs mass
regions up to 1 TeV. At hadron colliders the main Higgs production mechanism
is the gluon fusion \cite{H2gQCD0}, a process whose knowledge is fundamental
in order to  put limits on the Higgs mass or, in case the Higgs is discovered,
to compare the measured cross section with the SM result.
Concerning the Higgs decay channels,
it is quite difficult for an hadron collider to access
part of the mass range favored by the LEP results, 
the so called intermediate Higgs mass region $114 \lesssim 
\mh \lesssim 160$ GeV, 
because of the large QCD background to the dominant modes. In this
region the rare decay $ H \rightarrow \gamma \, \gamma $ is the most
interesting alternative to the usual decay channels.

It is then natural to try to have a theoretical prediction of the 
gluon fusion production cross section as well as this decay mode, as precise 
as possible. In the recent years most of the attention to these processes
has been devoted to the calculation of the QCD corrections, but we have
now reached a stage such that also the electroweak effects can be interesting.
In this report we make a first step in the calculation of the electroweak 
corrections by presenting the analytic result for the two-loop contributions
induced by the light (assumed massless) fermions. This subset of corrections, 
finite and gauge invariant, is actually  interesting for two reasons:
i) it could be numerically not irrelevant because one has to sum over
the generations; ii) because of the presence in the diagrams of two mass 
scales, $\mh$ and $\mw$, that in the most physically interesting Higgs mass
range are relatively close, and of cuts connected to the zero mass fermions, 
this subset cannot be computed 
with the standard heavy mass or momentum expansion method
but requires a complete calculation, making it a good probe of new techniques
for the computations of two-loop integrals.

We start analyzing the production cross section $\sigma(gg\to H)$.
Because of its importance this process has been
investigated with great accuracy in the recent years and it is now known at 
the  next-to-next-to-leading order in 
QCD \cite{H2gQCD1,QCDg2,H2gQCD2}. A recent discussion \cite{bd4}
on the residual theoretical uncertainty from perturbative QCD contributions
estimate it to be below 10\% for $\mh < 200 $ GeV. Electroweak corrections to
this production mechanism were only considered in the large-$\mt$ limit
and found to give a very small effect, below 1\% \cite{DjG}.

The Higgs boson, carrying no color, has no tree-level coupling to
gluons; therefore this process proceeds via loops.
At the partonic level the cross section, not corrected by QCD effects,
can be written as:
\bea
\sigma \left( g\,g \rightarrow  H  \right) & = &
\frac{G_\mu \alpha_s^2 }{512\, \sqrt{2} \, \pi} 
\left| {\cal G} \right|^2 \, ,
\label{ggh}
\eea 
where 
the lowest order one-loop  contribution is only due to the top quark and is 
given by:
\bea
 {\cal G}^{1l}_t &=& - 4 \,\th \, \left[ 2 -  (1 - 4 \,\th)\,
   H \left( -r,-r; -\frac1{\th} \right) \right] ~,
\label{eq:onelooptop}
\eea
with $\th \equiv \mt^2/\mh^2$ and\footnote{All the 
analytic continuations are obtained with the replacement 
$x \rightarrow x -i\,\epsilon$}
\be
 H (-r,-r; x ) = \frac12 
\log^2 \left( \frac{\sqrt{x+4}-\sqrt{x}}{\sqrt{x+4}+\sqrt{x}} 
\right)~.
\label{eq:C0}
\ee
At one-loop the contribution of light fermions is suppressed 
by their coupling to the Higgs  and it is completely negligible.

\begin{figure}[t]
\centering
\epsfig{file=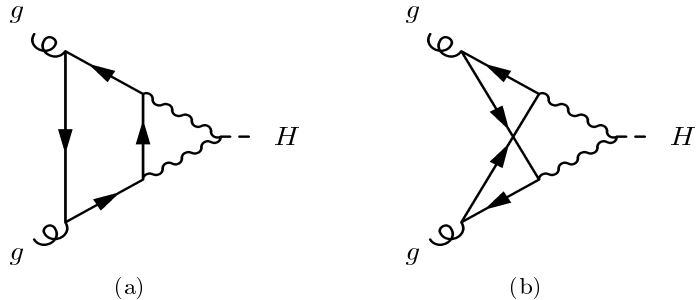,height=3.9cm,width=5.cm,
        bbllx=225pt,bblly=633pt,bburx=375pt,bbury=763pt}
\caption{Two-loop topologies involving light fermions contributing
to $gg \rightarrow H$.}
\label{fig1}
\end{figure}

At the two-loop level the light fermions can contribute to this process
because their  Higgs coupling suppression can be avoided by  coupling them 
to the $W$ or $Z$ bosons that can directly couple to the Higgs particle. 
Indeed the  light fermions contribute to the Higgs production through the 
topologies $(a)$ and $(b)$ depicted in Fig.~\ref{fig1}, with the $W$ or $Z$ 
boson exchanged in the loops together with quarks.

The general structure of the amplitude for the production of a Higgs particle,
in the fusion process of two gluons
of polarization vectors $\epsilon_\mu (q_1)$ and $\epsilon_\nu (q_2)$,
can be written as:
\be
T^{\mu \nu} = q_1^\mu \, q_1^\nu\, T_1 + q_2^\mu \, q_2^\nu\, T_2 +
              q_1^\mu \, q_2^\nu\, T_3 + q_1^\nu \, q_2^\mu\, T_4 +
              (q_1\cdot q_2) \,g^{\mu \nu} \,T_5 + 
         \epsilon^{\mu \nu \rho \sigma} \,q_{1 \rho} \,q_{2 \sigma}\, T_6 \, ,
\ee
with gauge invariance requiring that $T_1 = T_2 =0$ and $T_4 = -T_5$
while $T_3$ does not contribute for on-shell gluons and $T_6$ vanishes
exactly.
The relevant  form factors $T_4$ and $T_5$  are extracted from the 
one-particle irreducible (1PI) diagrams of Fig.~\ref{fig1} with the 
use of standard  projectors. The evaluation of  the 1PI part of the 
form factors is performed in few steps. First, the scalar amplitudes 
are reduced to a set  of linearly independent ones. This step is achieved 
in two different ways \cite{AB1}: i) a reduction in terms of scalar 
integrals containing only denominators is obtained with the introduction of 
a fictitious auxiliary planar double box diagram \cite{Glover:2000zu},  
with final momenta equal to the initial ones, that  contains an ``extra'' 
propagator with respect to the diagrams of Fig.~\ref{fig1}, such that the 
seven possible invariants that can appear in the numerator of a scalar 
amplitude can be simplified against the seven propagators of the auxiliary 
diagram. ii) By shifting  the loop momenta in an appropriate way we write 
the scalar amplitudes  in terms of  integrals with a number of propagators 
equal or smaller than the original one, having a set of independent
scalar products in the numerators \cite{B1}. Next, the independent scalar 
amplitudes obtained in the two reduction methods are expressed through the 
use of integration-by-parts  identities \cite{ibp} as a linear combination of 
the same  finite set of basic integrals, called master integrals (MIs).
The subsequent evaluation of the MIs is performed employing the technique
of the differential equations  in the external kinematical variables
\cite{dem}. Finally, the coefficients of the MIs Laurent expansion 
in $\epsilon = (4 -n)/2$, where $n$ is the dimension of the space-time,
are  expressed in closed analytic form. The 1PI diagrams of Fig.~\ref{fig1} 
show thresholds in the external  Higgs  momentum, $q^2$, at 
$q^2 =0,  \mw^2,  4 \,\mw^2$. Because of the presence of the
$q^2 = \mw^2$ and $q^2 = 4 \,\mw^2$ thresholds  our analytic result cannot be 
expressed only  in terms of Harmonic Polylogarithms (HPLs)~\cite{RV,GR}, 
a generalization of 
Nielsen's polylogarythms, but requires a recently introduced extension of the 
HPLs, the so-called Generalized Harmonic Polylogarithms (GHPLs)~\cite{AB2}. 
We explicitly checked that our analytic result satisfies all
gauge-invariance requests.

The total light fermion contribution to the partonic production cross section 
$\sigma(g g \rightarrow  H)$ is obtained summing over the first two
generations 
of quarks both the diagrams with the $W$ and those with $Z$ bosons. For the 
bottom quark instead, we can include only the contribution due to the $Z$ 
boson, which is by itself finite and gauge invariant.
We find, in the same normalization 
of Eq.~(\ref{ggh}) and in units $\alpha/(2 \pi s^2) (\mw^2/\mh^2)$:
\be
{\cal G}^{2l}_{lf} =  
  \frac{2} {c^4} \left(\frac54 - \frac73\, s^2 +
\frac{22}{9} s^4 \right)  A_1 \left[\zh \right] + 
4 \, A_1 \left[\wh \right] \, ,
\ee
where $\wh \equiv \mw^2/\mh^2$, $\zh \equiv \mz^2/\mh^2$,
$s^2 \equiv \sin^2 \theta_W$, $c^2 = 1- s^2$ and
\bea
 A_1 [x] & = &
   -4 + \! 2 \left( 1 - \Mvariable{x} \right)  
       H \! \left(\! -1;- \frac{1}{\Mvariable{x}} \right) - 
      2 \, \Mvariable{x} \, H \left(\! 0,-1;
        - \frac{1}{\Mvariable{x}} \right) \! + \! 
      2 \left( 1 - 3 \, \Mvariable{x} \right)  
       H \left( \! 0,0,-1;- \frac{1}{\Mvariable{x}} \! \right) \nn \\
&& +   2\,\left( 1 - 2\,\Mvariable{x} \right) \,
       H \left(0,-r,-r;- \frac{1}{\Mvariable{x}} \right) - 
      3\,\left( 1 - 2\,\Mvariable{x} \right) \,
       H \left(-r,-r,-1;- \frac{1}{\Mvariable{x}} \right) \nn \\
&& - \,  {\sqrt{ 1- 4\,\Mvariable{x}}}\,
       \left[ 2\,H \left(-r;-\frac{1}{\Mvariable{x}} \right) -
         3\,\left( 1 - 2\,\Mvariable{x} \right) \,
          H \left(-4,-r,-1;- \frac{1}{\Mvariable{x}} \right) \right. \nn \\
&&\left.  +  2 \left( 1 - 2 \, \Mvariable{x} \right)  
          H \left(-r,0,-1;- \frac{1}{\Mvariable{x}} \right) +  
         2 \left(1 - 2 \, \Mvariable{x} \right)  
          H \left(-r,-r,-r;- \frac{1}{\Mvariable{x}} \right) \right] ~.
\label{eq:Zcont}
\eea
Eq.~(\ref{eq:Zcont}) is expressed in terms of ordinary HPLs, the $H$ functions 
with only indices $0, -1$, and of GHPLs, $H$ functions where also the indices 
$-r$ and $-4$ are present. Their explicit expressions are presented in 
Appendix A.

\begin{figure}
\bc
\includegraphics[height=80mm,angle=0]{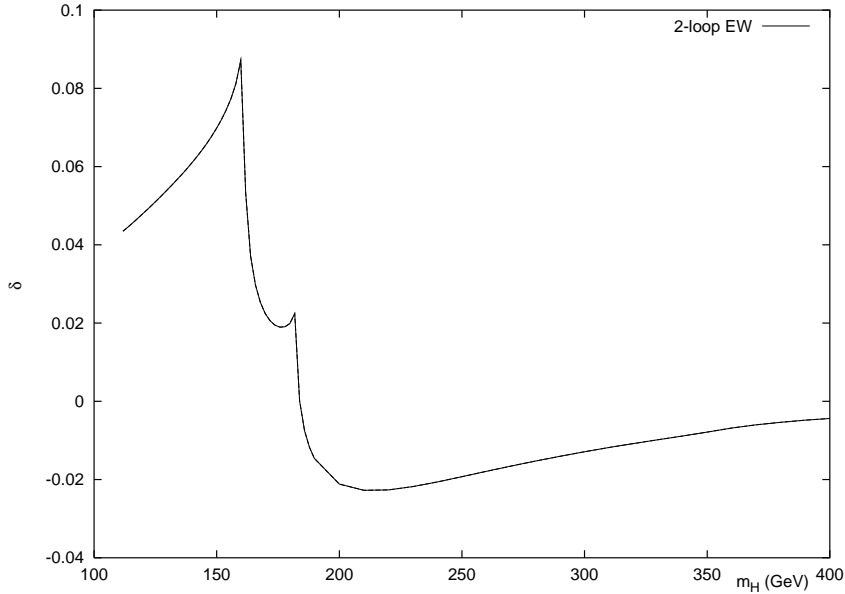}
\caption{\label{fig2} Relative corrections 
to the production cross section $\sigma( g g \rightarrow  H)$ and
to the decay width $\Gamma \left( H \rightarrow g g \right)$.}
\ec
\end{figure}
In Fig.~\ref{fig2} we  plot the relative corrections $\delta$, induced by 
the two-loop light fermion contribution, to the Higgs production cross section 
$\sigma \equiv \sigma_0 (1+\delta)$, where $\sigma_0$ is the lowest order 
result.
The correction raises till the opening of the $2\, W$ threshold in the 
topologies ({\em a, b}) of Fig.~\ref{fig1}, reaching up to 9\% of the one-loop 
result. As soon as the threshold is passed the correction has a sharp decrease 
with only a small 
bump at the opening of the $2 \, Z$ threshold, and for values of $\mh$ larger
than 200 GeV remains always smaller than 2\%. The large (for an electroweak 
correction)  result found in the intermediate Higgs mass region 
can be explained noticing that the topologies
({\em a, b}) of Fig.~\ref{fig1} with their thresholds at $2\,W$ and $2 \, Z$ 
are of a new type with respect to the 
one-loop result  that is described by a top loop triangle diagram.
We notice that, in the region below the $2 \,W$ threshold, the size of the 
electroweak light fermion corrections is 
comparable  to the accuracy  reached in the knowledge
of the QCD corrections to this production cross section \cite{bd4}.

The techniques used to compute the two-loop light fermion corrections
to the gluon fusion amplitude can be applied to study also the decay
processes $H \rightarrow \gamma\, \gamma,\,\,g\,g$.
The result for the latter is very easy to derive because one can
relate the partial decay width $\Gamma ( H \to gg)$ to 
$ \sigma \left( g\,g \rightarrow  H  \right) $ via:
\bea
\Gamma \left( H \rightarrow  g g  \right) & = &
\frac{8 \mh^3}{\pi^2} \, \sigma(g g\rightarrow H) \, .
\label{h2g}
\eea 
Thus our result for the light fermion corrections to the production
cross section can be translated into a result for the corrections to 
the partial decay width $\Gamma ( H \rightarrow g g)$ and Fig.~\ref{fig2}
can also be read as the relative corrections to this decay mode.
It should also be noticed that the QCD corrections to this decay are 
extremely large \cite{QCDg1,QCDg2}. As an example, in the intermediate 
Higgs mass region they shift the decay width by about 60-70\% upward.

We consider now the decay $ H \rightarrow \gamma \, \gamma$ that is of 
particular interest for the Higgs searches at hadron colliders
thanks to its clean signature. The Higgs boson, being a neutral particle, 
has no tree-level coupling to photons, therefore also this  decay 
can proceed only via loops. 
The corresponding partial decay width can be written as: 
\bea
\Gamma \left( H \rightarrow \gamma \, \gamma \right) &=& 
\frac{G_\mu \alpha^2 \, \mh^3}{128\, \sqrt{2} \, \pi^3} 
\left| {\cal F} \right|^2 \, ,
\label{eq:G}
\eea
where the lowest order one-loop contribution is due to $W$ boson and top loops,
and is given by \cite{oneloop}:
\bea
{\cal F}^{1l} &=& {\cal F}^{1l}_\smallw + \frac49 \,N_c \, {\cal F}^{1l}_t~, \\
{\cal F}^{1l}_\smallw &=& 2\, (1 + 6 \,\wh) - 12\,  \wh \,(1 - 2\, \wh)\,
H \left(-r,-r; -\frac1{\wh} \right)~,  \\
 {\cal F}^{1l}_t &=& {\cal G}^{1l}~,
\label{eq:oneloopw}
\eea
with  $N_c$ the color factor. 

The $W$ and top one-loop contributions interfere destructively. 
They approach constant values (${\cal F}^{1l}_\smallw \rightarrow 7, \,
{\cal F}^{1l}_t \rightarrow -4/3$) for  mass of the particle inside the loop
much heavier than $\mh$, with the $W$ loops  always providing the dominant
part for Higgs mass up to $\mh \sim 600$ GeV, where the top 
contribution becomes comparable.

Two-loop investigations of this decay channel have been mainly devoted to
QCD corrections to the one-loop top contribution \cite{QCD2loop}. Concerning 
two-loop electroweak corrections only the ${\cal O} (G_\mu \mt^2)$ 
term \cite{EW2lmt} and the ${\cal O} (G_\mu \mh^2)$ one \cite{EW2lmh}  were 
studied. 
\begin{figure}[t]
\centering
\epsfig{file=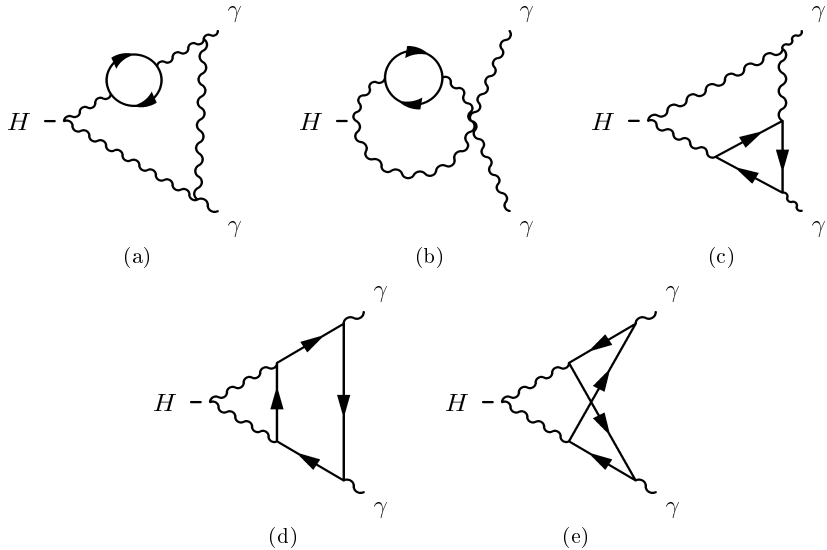,height=7cm,width=9.cm,
        bbllx=150pt,bblly=320pt,bburx=450pt,bbury=570pt}
\caption{Two-loop topologies involving light fermions contributing
to $H \rightarrow \gamma \gamma$.}
\label{fig3}
\end{figure}

Differently from the one-loop level at  two-loop the light fermions can 
contribute to this  partial decay width through all diagrams
whose topologies are depicted in Fig.~\ref{fig3}. 
The choice of an appropriate gauge-fixing procedure
can somewhat reduce the number of diagrams to be computed. To achieve 
this we find it convenient to employ the Background Field Method (BFM) 
quantization framework.
The BFM is a technique for quantizing gauge  theories \cite{BFM,abbott} that 
avoids the complete explicit breaking of the gauge symmetry.
One of the salient features of this approach is that all fields are
splitted in two components: a classical background field $\hat{V}$
and a quantum field $V$ that appears only in the loops.
The gauge-fixing procedure is achieved through a non linear term in the
fields that breaks the gauge invariance only of the quantum part of the
lagrangian, preserving the gauge symmetry of the effective action with respect 
to the background fields. Thus, the BFM gauge-fixing function induces a 
modification of some of the vertices in which  background fields are present, 
such that, in the BFM Feynman gauge ($\xi_Q =1$, $\xi_Q$ being the gauge 
parameter), the vertex coupling a background (external) photon to a quantum 
$W$ boson and its unphysical scalar counterpart is absent\footnote{A general 
discussion on the application of the BFM to the SM with the complete set of 
BFM Feynman rules can be found  in Ref.\cite{ddw} while a
specific two-loop calculation in this framework is presented in 
Ref.\cite{DV}.}. With this choice of gauge only the physical $W$ and $Z$ bosons
are exchanged in Fig.~\ref{fig3}. In particular, in the topologies 
({\em a, b, c}) the internal wavy line can only be a $W$ boson while in the
topologies ({\em d, e}) it can represent both a $W$ and a $Z$ boson.
The evaluation of the diagrams of  Fig.~\ref{fig3} follows the line
already discussed for the gluon fusion cross section.
Concerning the counterterm diagrams, 
which are not depicted in Fig.~\ref{fig3},
their evaluation does not present any particular difficulty. We just
notice that the contributions coming from the  ${\cal O}(\epsilon)$ part 
of the one-loop diagrams cancel out in the total result. The renormalization
is actually straightforward. Because we are in the presence of two real 
photons their coupling is given by $\alpha = 1/137.036$ while the coupling
of the Higgs to the vector boson is more appropriately described in terms of
$G_\mu$. Correspondingly, we employ the on-shell definition for the $W$ boson
mass.

The full light fermion contribution to the Higgs decay into
two photons is obtained summing all the diagrams over the 3 lepton
families plus 
the two first generations of quarks. Concerning the third generation, the 
bottom  contribution to topologies ({\em d, e}) mediated by the $Z$ boson, 
that is finite and gauge-invariant,  can also be included. 
The light fermion contribution to the partial width 
$H \rightarrow \gamma \, \gamma $,  in the same normalization of 
Eq.~(\ref{eq:G}), reads
in units  $\alpha/(2 \pi s^2) (\mw^2/\mh^2)$:
\bea
{\cal F}^{2l}_{lf} &=&  
2 \,N_c\, A_2 \left[-2/9, \wh\right] + 
3 \,A_2 \left[0,\wh \right] +
  \frac{2\,N_c} {c^4} \left(\frac{11}{36} - \frac{19}{27}\, s^2 +
\frac{70}{81} s^4 \right)  A_1 \left[\zh \right] \nn \\
&& + \frac3{c^4}\, \left(\frac12 -2\, s^2 + 4\, s^4 \right) 
 A_1 \left[\zh \right] \, ,
\label{eq:twoloop}
\eea 
where 
\bea 
 A_2 [q,x] & = &
     -8 \left( 1 +q \right)  + 
      4 \left( 1 +q \right)  
       \left( 1 - \Mvariable{x} \right)  
       H\left(-1;- \frac{1}{\Mvariable{x}}  \right) - 
      2 \left( 1 + 2 \, q \, \Mvariable{x} \
\right)  H\left(0,-1;-\frac{1}{\Mvariable{x}}\right) \nn \\
&& - \frac{2}{3} \left( 5 - 12 \, \Mvariable{x} \right)  
         H\left(\!-r,-r;- \frac{1}{\Mvariable{x}} \! \right) - 
      6 \left( 1 +q - 3  \Mvariable{x} -
         2 \, q \, \Mvariable{x} \right)  
       H\left(\!-r,-r,-1;- \frac{1}{\Mvariable{x}}\! \right)  \nn \\
&& +   2\,\left( 1 + 2\,q \right) \, \left[ 
       \left( 1 - 2\,\Mvariable{x} \right) \,
       H\left(0,-r,-r;- \frac{1}{\Mvariable{x}} \right) +
       \left( 1 - 3\,\Mvariable{x} \right) \,
       H\left(0,0,-1;-\frac{1}{\Mvariable{x}}  \right) \right]  \nn\\
&& -  {\sqrt{1-4\,\Mvariable{x}}}\,
       \left\{ 2\,\left( 1 + 2\,q \right) \,
          H\left(-r;- \frac{1}{\Mvariable{x}} \right) - 6\,q\,
          \left( 1 - 2\,\Mvariable{x} \right) \,
          H\left(-4,-r,-1;- \frac{1}{\Mvariable{x}} \right) \right.\nn \\
&&  ~~~~~~~~~~~~~~  \left.  + 4\,q\, \left( 1 - 2\,\Mvariable{x} \right) \,
       \left[   H\left(-r,0,-1;- \frac{1}{\Mvariable{x}}\right) 
        +  H\left(-r,-r,-r;- \frac{1}{\Mvariable{x}} \right) \right] 
\right\} \nn \\
&&  + \frac{6\,{\left( 1 - 2\,\Mvariable{x} \right) }^2}{ 
            \sqrt{1-4\,\Mvariable{x}}}\,
            H\left(-r,-1;-\frac{1}{\Mvariable{x}} \right)  \, ,
\label{eq:Wcont}
\eea
and $A_1$ has been given in Eq.~(\ref{eq:Zcont}).

Eq.~(\ref{eq:Wcont}) shows an unphysical singularity at $x = \frac14$, i.e. at 
the opening of the $2\, W $ threshold, connected to the appearance in its
last line of a term with a square root in the denominator that  exactly at the 
threshold becomes formally infinite. This unphysical infinity is actually a 
signal  that our first order treatment of  the $W$ propagator in topologies 
({\em a, b}) of Fig.~\ref{fig3}
is inadequate in the $2 \,W$  threshold region. To obtain a finite
result also in this region we  regulate the singularity by  performing  
the replacement $\mw \rightarrow \mw -i \Gamma_\smallw/2$ 
in the square root and we check the dependence of the final result on the
regulator.

\begin{figure}
\bc
\includegraphics[height=80mm,angle=0]{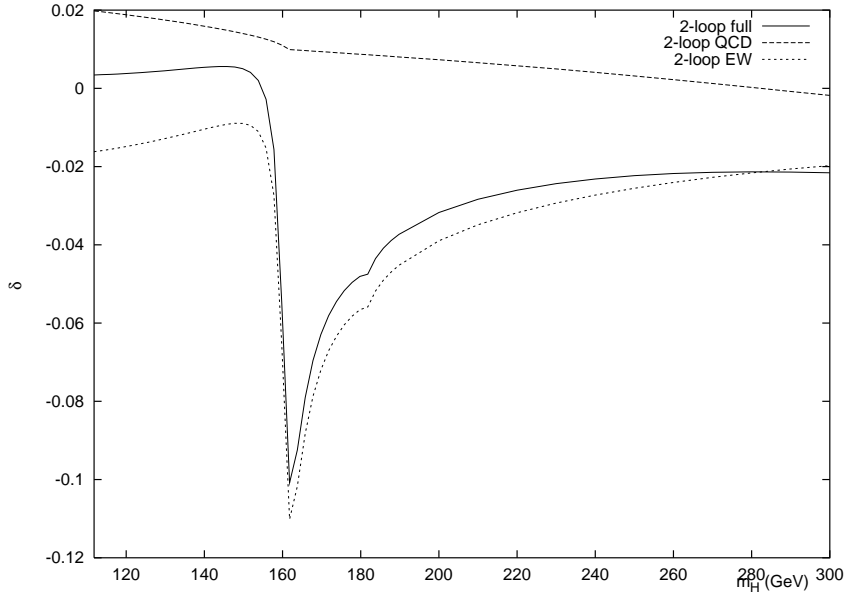}
\caption{\label{fig4}Relative corrections to the decay width 
        $H \rightarrow \gamma \gamma$.}
\ec
\end{figure}
In Fig.~\ref{fig4} we  plot separately the relative corrections, $\delta $, 
to the decay width $\Gamma \equiv \Gamma_{0} (1 + \delta)$ 
$( \Gamma_{0} \propto |{\cal F}^{1l}|^2)$ induced by the two-loop
light fermion contribution and by the QCD corrections to the one-loop top
term as well as the sum of the two contributions. The figure is obtained 
by regulating the singularity at $x = \frac14$ with 
$\Gamma_\smallw/\mw = 2.5 \cdot 10^{-2}$. We check the dependence of the result
on the regulator by varying  $\Gamma_\smallw/\mw$ down to  $2.5 \cdot 10^{-9}$.
Above the threshold already at $\mh = 165$ GeV the result changes by less 
than 5\% and becomes insensitive to the size of the regulator 
at $\mh = 170$ GeV. Below the threshold,
down to $\mh = 150$ GeV the variation is larger. However, for a
Higgs mass value around 150 GeV the light fermion correction is 
small and therefore it is not surprising that the relative variations
are large. It is clear that in the region $150 \lesssim \mh \lesssim 170$ GeV 
our result should be taken with some caution.

As it can be seen from
Fig.~\ref{fig4} the QCD and light fermion  contributions for 
$\mh \lesssim 140$ GeV  are of comparable size  but with opposite sign
so that the total correction is below 1\%.
As an example, for $\mh = 120$ GeV the light fermion contribution
amounts to $\sim$($-$1.5\%) of the total width  while the QCD one gives 
$\sim$($+$1.8\%). For values of $\mh$ above 140 GeV the light fermion 
correction becomes   small until the region close to the opening of the
$2\, W$ threshold is reached. Crossing the threshold the correction becomes  
quite large, in absolute value, but as soon as the $2\, W$ threshold region is 
passed it starts to decrease again. However, for $\mh$ around 200 GeV, we find
still a $\sim$5\% correction while the QCD contribution 
is relatively small.  The visible bump in the figure around 180 GeV
is actually connected with the opening of the $2\, Z$ threshold in the 
topologies ({\em d, e})  of Fig.~\ref{fig3}.

In conclusion, we have computed the two-loop electroweak corrections induced
by the light fermions to the Higgs gluon fusion production cross section 
$\sigma ( g g \rightarrow H)$ and to the partial decay widths 
$\Gamma \left( H \rightarrow  \gamma \, \gamma  \right)$ and 
$\Gamma \left( H \rightarrow  g \, g \right)$. 
For the gluon fusion production cross section we have found that the light 
fermion corrections are quite large in the region below the $2 \, W$ threshold
reaching up to 9\% of the lowest order result. In the case of the Higgs decay 
to two photons, the corrections are  large above the $2 \,W$ threshold,
while below it they are quite small a part the region $\mh \lesssim 140$ GeV 
where they are larger than 1\% cancelling part of the  QCD corrections 
giving a  total correction below 1\%. It should be recalled that, both for the 
production and the decay,  the light fermions are only a small part of the 
electroweak corrections. A complete  calculation of the electroweak 
corrections to 
$\sigma ( g g \rightarrow H)$ seems to be relevant to assure the accuracy
of the knowledge of this important production cross section. 
\vspace{0.5cm}

One us (G.D.) wants to thank F.~Maltoni for interesting discussions. 
This work was partially supported by the European Community's
Human Potential Programme under contract
HPRN-CT-2000-00149 (Physics at Colliders).

\appendix
\section{Generalized Harmonic Polylogarithms}
In this Appendix,  we outline the definitions and properties of the 
HPLs and GHPLs and give the explicit expressions of those entering in
Eqs. (\ref{eq:Zcont},\ref{eq:Wcont}).

The idea lying behind the introduction of HPLs is to express a given 
integral coming from the calculation of a Feynman diagram in a unique 
and non-redundant way as a linear combination of a minimal set of 
independent transcendental functions. These functions are expressed as 
repeated integrations over a starting set of basis functions and this
set depends strongly on the problem one has to solve, being connected 
directly to the threshold structure of the diagrams under consideration. 
In the case in which only thresholds at $0$ and $4m^2$ or at $0$ and $m^2$ 
occur, the original set of HPLs  introduced in 
\cite{RV} is sufficient to cover all the possible integrals. However, 
for different structure of the thresholds of the Feynman diagrams involved 
in the calculation, as it is the case at hand, one is forced to enlarge 
the set of starting basis functions.

In our case, we  enlarge the original basis functions 
with indices $\{ -1,0,1 \}$:
\be
g(-1;x) = \frac{1}{1+x} \, , \quad  g(0;x) = \frac{1}{x} \, , \quad
g(1;x) = \frac{1}{1-x} \, ;
\ee
that give a ``weight-1'' HPLs:
\be
H(-1;x) = - \log (1-x) \, , \quad H(0;x) = \log x \, , \quad
H(1;x) = \log (1+x) \, ,
\label{peso1} 
\ee
with the following two basis functions \cite{AB2}:
\be
g(-4;x) = \frac{1}{4 + x} \, , \quad g(- r;x) = 
\frac{1}{\sqrt{x(4 + x)}} \, ,
\ee
with the corresponding weight-1 GHPLs:
\be
H(-4;x) = \log (4+x) - 2 \log 2 \, , \quad 
H(-r;x) = - \log z \, ,
\label{peso1a} 
\ee
where:
\be
z = \frac{\sqrt{x+4}-\sqrt{x}}{\sqrt{x+4}+\sqrt{x}} \, .
\label{chvar}
\ee
Integrals with weight $>1$ are defined using the
recursive definitions:
\be
H(\vec{0}_{w};x) = \frac{1}{w\!} \log^{w}{x} \, , \quad
H(a,\vec{w};x) = \int_{0}^{x} dt \, g(a;t)  H(\vec{w};t) \, , 
\label{wn2} 
\ee
where $\vec{0}_{w}=(0,0, \cdots ,0)$ is a vector containing $w$ zeroes,
$\vec{w}$ is a vector with $w$ components running over the set 
$\{ -4,-r,-1,0,1 \}$ and $a$ takes the values $-4,\,-r,\, -1,\,0,\, 1$.

It has been shown \cite{RV,GR,AB2} that the set of (G)HPLs constructed in 
this way satisfies useful relation among themselves. As an example,
the product of two GHPLs of weight $w_1$ and $w_2$ can be 
expressed as a linear combination of GHPLs of weight 
$w_1 + w_2$.

Eqs. (\ref{eq:Zcont},\ref{eq:Wcont}) contain 11 GHPLs. 
$H(-r,-r;x)$ is defined in Eq.~(\ref{eq:C0}) while 
$H(-1;x)$ and $H(-r;x)$ have already been 
presented in  Eqs. (\ref{peso1},\ref{peso1a}); the other 8 are:
\bea
H(0,-1;x) & = & - {\rm Li}_2 (-x) \, , \\
H(0,0,-1;x) & = & - {\rm Li}_3 (-x) \, ,\\
H(-r,-r,-r;x) & = & - \frac{1}{6}  \log^{3}{z} \, , \\
H(0,-r,-r;x) & = & - \frac{1}{6}  \log^{3}{z} + 2\, {\rm
S}_{1,2}{(1-z)} \, \\
H(-r,-1;x) & = & \int_{0}^{x} \frac{dt}{\sqrt{t(4+t)}} \log{(1+t)} \, , \\
H(-r,0,-1;x) & = & \int_{0}^{x} \frac{dt}{\sqrt{t(4+t)}} 
\int_{0}^{t} \frac{dq}{q} \log{(1+q)} \, , \\
H(-r,-r,-1;x) & = & \int_{0}^{x} \frac{dt}{\sqrt{t(4+t)}} 
\int_{0}^{t} \frac{dq}{\sqrt{q(4+q)}} \log{(1+q)} \, , \\
H(-4,-r,-1;x) & = & \int_{0}^{x} \frac{dt}{(4+t)}  
\int_{0}^{t} \frac{dq}{\sqrt{q(4+q)}} \log{(1+q)} \, .
\eea
where ${\rm{Li_2}} (x) = - \int_0^x dt \, {\log (1-t) \over t} $,
and ${\rm{Li_3}} (x) = \int_0^x dt \, { {\rm{Li_2}} (t) \over t}$ 
are the Nielsen's polylogarithms and
${\rm S}_{1,2}{(1-z)}=\frac{1}{2} \int_{0}^{1-z}dt \, \frac{\log^{2}{(1-t)}}{t}$ 
is a Spence function.

These 11 GHPLs are real and analytic in all the half plane $x \geq 0$ and
they present cuts in the plain $x<0$. In our case the variable $x$ is 
defined as $x=-s/m^2$, 
where $s$ is the squared c.m. energy and $m$ is the mass of the particle
running in the loop ($W$ or $Z$ bosons). In the physical region 
$s=m_H^2>0$, the variable $x$ is negative, and the GHPLs have to be 
continued analytically. The analytical continuation is done with the 
usual $i \epsilon$-prescription, i.e. giving a small positive imaginary 
part to $s$: $s+i \epsilon$. In so doing, we have: 
\be
x \rightarrow -y -i \epsilon \, ,
\ee
where $y=s/m^2 >0$, and, for example:
\bea
\sqrt{x} & \rightarrow & \sqrt{-y-i \epsilon} = -i \sqrt{y} \, , 
\label{ac1} \\
\log{x} & \rightarrow & \log{(-y-i \epsilon)} = \log{y} - i \pi \, .
\label{ac2}
\eea
The numerical evaluation of the 11 GHPLs comes straightforwardly from their
definition as repeated integrals.

\end{document}